\documentclass[12pt]{article}

\newcommand{\bea}{\begin{eqnarray}}
\newcommand{\eea}{\end{eqnarray}}

\title{
\textbf{On conformal invariant integrals involving spin one-half and spin-one
particles}}
\author{
Indrajit Mitra\footnote{E-mail: imitra@iitk.ac.in}\\\\
Department of Physics, Indian Institute of Technology,\\ 
Kanpur 208016, India}
\date{}
\begin{document}
\maketitle
\begin{abstract}
We consider the evaluation of $D$-dimensional conformal invariant 
integrals which involve spin one-half and spin-one particles. The 
star-triangle relation for the massless Yukawa theory is derived, and the 
longitudinal part of the three-point Green function
of massless QED is determined to the lowest order in position space. 
The operator algebraic method of calculating massless Feynman 
integrals is used for the evaluation.
\end{abstract}

\section{Introduction}
\label{intro}
This work is concerned with the evaluation of conformal invariant integrals
in Euclidean space with general number of dimensions. A scale and Poincare invariant field theory
is generally also conformal invariant. Now, any theory of massless
particles with dimensionless couplings is scale invariant at the tree level.
Therefore, tree-level integrals in position space in such a theory will exhibit a
conformal invariant structure. The simplest example of this is the star-triangle relation
(also called the uniqueness relation) involving three massless scalar fields. 
This relation, which evaluates the integral for the tree-level three-point function,
not only brings out the
conformal structure, 
but also evaluates the coefficient exactly. The star-triangle relation
in three dimensions was given in Ref.\ \cite{deramo}, and was proved for general number
of dimensions by Symanzik in
Ref.\ \cite{sym}.

Now, the conformal invariance of the tree level can get broken due to diverging loop
contributions. Even then, such exact relations at the tree level are useful for carrying out the
integration over the internal vertices in position space diagrams. Of particular importance,
however, is the application of such relations to conformal field theories (CFT's).
Various aspects of CFT's in $D$ dimensions have been reviewed in Refs.\ \cite{fradkin1}
and \cite{fradkin2}. The evaluation of the tree level integrals is necessary for implementing
the bootstrap program in a CFT \cite{migdal, parisi}. 
Using the star-triangle relation to integrate
over the internal vertices, the bootstrap program
has been carried out to determine the anomalous dimensions in $\phi^4$ theory \cite{ano}.
However, $D$-dimensional CFT's generally involve 
particles with non-zero spin. A well-known example is the ${\cal N}=4$ supersymmetric 
Yang-Mills theory. Calculation of Feynman integrals in the position space
in this theory have been carried out in several recent works \cite{cvetic}. 
In the context of massless QED, a formulation 
of conformal QED$_4$ was suggested in Refs.\
\cite{palchik} and \cite{comm}. Also, the infrared limit of massless  QED$_3$ is a CFT
\cite{mrs1, mrs2}. 
Possible application in these theories is a motivation for studying conformal 
invariant integrals with
spinors and vector particles.

Other than this, analytical evaluation of massless Feynman integrals at multi-loop level and the
star-triangle relation are generally important for calculations in perturbative field
theory at high orders, and in mathematical physics: see Ref.\ \cite{isaev1} and references
therein. Recently, a simple 
method of doing these calculations have developed by Isaev
\cite{isaev1, isaev2} which replaces the Feynman integrals by algebraic manipulation
of operators. We use this method extensively in this paper.

Three-point functions involving conserved vector operators in $D$-dimensional CFT's have
been discussed in Ref.\ \cite{os}. We will, however, be concerned with the three-point function
involving the fermion and the gauge field in QED. This has been discussed in Refs.\ 
\cite{palchik} and \cite{comm}, and we will compare our result with that given in these two works.

The paper is organized as follows. In Sec.\ \ref{star}, we discuss the usual star-triangle
relation using the operator approach. In Sec.\ \ref{yukawa}, we derive the
star-triangle relation for the massless Yukawa theory. 
In Sec.\ \ref{qed},
we perform an explicit calculation of the
longitudinal part of the three-point Green function
of massless QED to the lowest order in position space. 
Our conclusions are presented in Sec.\ \ref{concl}.
\section{Star-triangle relation involving scalar fields}
\label{star}
It will be helpful to first discuss, following Ref.\ \cite{isaev1}, 
the usual star-triangle relation
involving scalar fields within the framework of the operator algebraic method. 
The relation evaluates $\langle 0|T(\phi_1(x_1)\phi_2(x_2)\phi_3(x_3))|0\rangle$ 
to the lowest order in position space with a $\phi_1\phi_2\phi_3$ interaction. With 
$x_{ab}\equiv x_a-x_b$, the relation is given by \cite{sym}
\bea
\int d^Dx_4~ (x_{14}^2)^{-\delta_1} (x_{24}^2)^{-\delta_2} (x_{34}^2)^{-\delta_3}
&=&\pi^{D/2}\frac{ \Gamma(D/2-\delta_1) \Gamma(D/2-\delta_2) \Gamma(D/2-\delta_3)}{\Gamma(\delta_1)\Gamma(\delta_2)\Gamma(\delta_3)}\nonumber\\
&&\times (x_{12}^2)^{-D/2+\delta_3} 
 (x_{13}^2)^{-D/2+\delta_2} (x_{23}^2)^{-D/2+\delta_1}\,,          \label{st}
\eea
where 
\bea
\delta_1+\delta_2+\delta_3=D\,.                                        \label{delta}
\eea
The left-hand side of Eq.\ (\ref{st}) represents the propagation of a massless scalar particle
between the point $x_a$ and the internal vertex $x_4$ with a scale dimension
$\delta_a$, for $a=1,2,3$.  It is to be noted that 
Eq.\ (\ref{delta}) ensures that the coupling constant of the
 $\phi_1\phi_2\phi_3$ interaction is dimensionless, and that the right hand side of
Eq.\ (\ref{st}) has the conformal structure of the three-point function involving three 
scalar fields also because of Eq.\ (\ref{delta}).

In the operator approach, one reduces Feynman integrals to products
of position and momentum operators $\hat{q}_i$ and $\hat{p}_i$ ($i=1,\cdots D$) taken between
position eigenstates. A collection of useful formulas
are given in the Appendix of our paper. The {\it key relation} 
(Eq.\ (9) of Ref. \cite{isaev1}) is
\bea
\hat{p}^{\,-2\alpha}\hat{q}^{\,-2(\alpha+\beta)}\hat{p}^{\,-2\beta}
=\hat{q}^{\,-2\beta}\hat{p}^{\,-2(\alpha+\beta)}\hat{q}^{\,-2\alpha}\,.              \label{key1}
\eea
This is {\it the star-triangle relation in the operator form}. To see this, one has to
take Eq.\ (\ref{key1}) between the states $\langle x|$ and $|y\rangle$. This gives, 
on inserting the completeness relation
and using Eqs.\ (\ref{eq:A4}), (\ref{eq:A5}) and (\ref{eq:A6}),
\bea
\int d^Dz~ \frac{1}{|x-z|^{D-2\alpha}}\frac{1}{|z|^{2(\alpha+\beta)}}
         \frac{1}{|y-z|^{D-2\beta}}
         &=&\pi^{D/2}\frac{\Gamma(\alpha)\Gamma(\beta)\Gamma(D/2-\alpha-\beta)}
                        {\Gamma(\alpha+\beta)\Gamma(D/2-\alpha)\Gamma(D/2-\beta)}\nonumber\\
        &&\times   \frac{1}{|x|^{2\beta}}\frac{1}{|x-y|^{D-2\alpha-2\beta}}
            \frac{1}{|y|^{2\alpha}}\,.                                         \label{key2}
\eea
{\it It is important to note from Eqs.\ (\ref{key1}) and (\ref{key2}) that the}
$``\hat{p}\hat{q}\hat{p}"$ {\it form represents the integral, while the}
$``\hat{q}\hat{p}\hat{q}"$ {\it form gives the result of integration}. Now let 
$x=x_1-x_2$ and  $y=x_3-x_2$, and let us also change to a new integration
variable $x_4$ defined by $z=x_4-x_2$. Also, define $\delta_1$, $\delta_2$ and $\delta_3$ by
$D/2-\alpha=\delta_1$,  $\alpha+\beta=\delta_2$ and $D/2-\beta=\delta_3$.
This leads us to the relation stated in the form of Eq.\ (\ref{st}). 
\section{Star-triangle relation for massless Yukawa theory}
\label{yukawa}
We now turn to the massless Yukawa theory with a $\bar\psi\psi\phi$ interaction and 
a dimensionless coupling.
Ref.\ \cite{sym} gives the method of
deriving the star-triangle relation for this theory using Schwinger parameters.
We show how the operator approach
provides us with an alternative method, and derive the relation. The manipulations which
we perform also set the
stage for the calculation of Sec.\ \ref{qed}.

The suitable starting $``\hat{p}\hat{q}\hat{p}"$ form is now
\bea
\Gamma\equiv\gamma_i\gamma_j \hat{p}_i\hat{p}^{\,-2\alpha-1}\hat{q}_j
\hat{q}^{\,-2(\alpha+\beta)-1}
\hat{p}^{\,-2\beta}\,,                                 \label{yu1}
\eea
so that
\bea
\langle x|\Gamma|y\rangle=i(D-2\alpha-1) a(\alpha+1/2)a(\beta)
\int d^Dz~\frac{\rlap/x-\rlap/z}{|x-z|^{D-2\alpha+1}}\,
            \frac{\rlap/z}{|z|^{2(\alpha+\beta)+1}}\,
            \frac{1}{|y-z|^{D-2\beta}}\,.                                   \label{yu2}
\eea
To convert $\Gamma$ to $``\hat{q}\hat{p}\hat{q}"$ form, we first
put $\hat{q}_j$ next to $\hat{p}_i$ in Eq.\ (\ref{yu1}) by using
Eq.\ (\ref{eq:A2}). (We extend the validity of 
Eq.\ (\ref{eq:A2}) to all real $\alpha$, and use it with
$-2\alpha-1$ in the place of $2\alpha$.) We thus obtain
\bea
\Gamma=\gamma_i\gamma_j \hat{p}_i\,(\hat{q}_j\hat{p}^{\,-2\alpha-1}
                                  +i(2\alpha+1)\hat{p}^{\,-2\alpha-3}\hat{p}_j)\,
                                 \hat{q}^{\,-2(\alpha+\beta)-1}\hat{p}^{\,-2\beta}\,.
\eea
Since $\gamma_i\gamma_j \hat{p}_i\hat{p}_j=\hat{p}^2$, we can use the key relation of
Eq.\ (\ref{key1}) (with $2\alpha+1$ in the place of  $2\alpha$) 
in both the terms and obtain
\bea
\Gamma=\gamma_i\gamma_j \hat{p}_i\hat{q}_j\hat{q}^{\,-2\beta}\hat{p}^{\,-2(\alpha+\beta)-1}
\hat{q}^{\,-2\alpha-1}
+i(2\alpha+1)\hat{q}^{\,-2\beta}\hat{p}^{\,-2(\alpha+\beta)-1}
\hat{q}^{\,-2\alpha-1}\,.                                                    \label{semi}
\eea
The second term is already of $``\hat{q}\hat{p}\hat{q}"$ form. To put the first term also
into that form, $\hat{p}_i$ has to be brought next to $\hat{p}^{\,-2(\alpha+\beta)-1}$.
For this, we take $\hat{p}_i$ first through $\hat{q}_j$ using Eq.\ (\ref{eq:A1})
and then through $\hat{q}^{\,-2\beta}$ using Eq.\ (\ref{eq:A3}). This generates a couple
of additional terms, and on simplification Eq.\ (\ref{semi}) reduces to
\bea
\Gamma=\gamma_i\gamma_j\hat{q}_j\hat{q}^{\,-2\beta}\hat{p}_i\hat{p}^{\,-2(\alpha+\beta)-1}
\hat{q}^{\,-2\alpha-1}
-i(D-2\alpha-2\beta-1)\hat{q}^{\,-2\beta}\hat{p}^{\,-2(\alpha+\beta)-1}
\hat{q}^{\,-2\alpha-1}\,.
\eea
To obtain $\langle x|\Gamma|y\rangle$, we use Eqs.\ (\ref{eq:A4}), (\ref{eq:A5}) and (\ref{eq:A7}).
This gives
\bea
\langle x|\Gamma|y\rangle=i(D-2\alpha-2\beta-1)a(\alpha+\beta+1/2)
              \frac{(\rlap/x-\rlap/y)\rlap/y}
                   {|x|^{2\beta}|x-y|^{D-2\alpha-2\beta+1}|y|^{2\alpha+1}}\,.    \label{yu3}
\eea
Equating the right-hand sides of Eqs.\ (\ref{yu2}) and (\ref{yu3}), we arrive at the desired
relation. The coefficients can be determined from Eq.\ (\ref{eq:A6}) and simplified using
the relation $n\Gamma(n)=\Gamma(n+1)$. Finally, using the new variables
$x_a$ and $\delta_a$ as below Eq.\ (\ref{key2}), we arrive at the
form
\bea   
&&\int d^Dx_4~ \frac{\rlap/x_{14}}{(x_{14}^2)^{\delta_1+1/2}}\,
              \frac{\rlap/x_{42}}{(x_{24}^2)^{\delta_2+1/2}}
               \frac{1}{(x_{34}^2)^{\delta_3}}\nonumber\\
&=&\pi^{D/2}\frac{ \Gamma(D/2-\delta_1+1/2) \Gamma(D/2-\delta_2+1/2) \Gamma(D/2-\delta_3)}
{\Gamma(\delta_1+1/2)\Gamma(\delta_2+1/2)\Gamma(\delta_3)}\nonumber\\
&&\times \frac{\rlap/x_{13}}{(x_{13}^2)^{D/2-\delta_2+1/2}}\,
          \frac{\rlap/x_{32}}{(x_{23}^2)^{D/2-\delta_1+1/2}}         
           \frac{1}{(x_{12}^2)^{D/2-\delta_3}}\,                                   \label{yu4}
\eea
As before, Eq.\ (\ref{delta}) ensures scale
invariance at the tree level. For the special case
of $D=4$, Eq.\ (\ref{yu4}) is in agreement with Eq.\ (A6.12a) of Ref.\ \cite{fradkin1}.
\section{Three-point Green function for massless QED}
\label{qed}
In order to evaluate $\langle 0|T(\psi(x_1)\bar\psi(x_2)A_k(x_3)|0\rangle$ at the lowest order,
we first need to specify the tree-level propagators in the position space.
We follow the convention \cite{fradkin1, fradkin2} of writing the behaviour of the fermion
propagator and the photon propagator as 
\bea
S(x)\sim\frac{\rlap/x}{(x^2)^{d_\psi+1/2}}\,,~~~
D_{kl}(x)\sim
\Bigg(\delta_{kl}-(1-\eta)\frac{\partial_k\partial_l}{\partial^2}\Bigg)
\frac{1}{(x^2)^{d_A}}\,.
\eea
Here $d_\psi$ and $d_A$ are the scale dimensions,
and $\eta$ is the gauge-fixing parameter.
The fermion propagator $\rlap/p/p^2$ in momentum space implies $d_\psi=(D-1)/2$.
For the photon, we will consider $d_A=1$ \cite{palchik, comm}.   
A photon propagator thus behaving as
$1/p^{D-2}$ in momentum space ensures that the QED coupling constant is dimensionless.
This behaviour, present in QED$_4$,
also occurs in massless QED$_3$ in the infrared: in the
latter theory, the photon propagator goes as $1/p$ in the infrared in the $1/N$
expansion ($N$ being the number of fermion flavours) \cite{appel, mrs1}.

The starting $``\hat{p}\hat{q}\hat{p}"$ form for the
lowest-order three point function is therefore
\bea
\hat{p}_i\gamma_i\hat{p}^{\,-2}\gamma_l \hat{q}_j\gamma_j\hat{q}^{\,-D}
                      \hat{p}^{\,-D+2}(\delta_{kl}-(1-\eta)\hat{p}_k\hat{p}_l\hat{p}^{\,-2})\,.                                                                                 \label{qed0}
\eea
(It may be helpful to compare Eq.\ (\ref{qed0}) with  Eq.\ (\ref{yu1}). In Eq.\ (\ref{qed0}),
we have $\alpha=1/2$ and $\beta=(D-2)/2$. There is also
a $\gamma_l$ vertex factor and the tensor part of the photon propagator.)
In the present work, we will consider only the {\it longitudinal} part:
\bea
\Gamma_k\equiv\eta\gamma_i\gamma_l\gamma_j\hat{p}_i\hat{p}^{\,-2}\hat{q}_j\hat{q}^{\,-D}
                      \hat{p}^{\,-D+2}\hat{p}_k\hat{p}_l\hat{p}^{\,-2}\,.         \label{qed1}
\eea
On using the position space ``matrix elements" of
$\hat{p}_i\hat{p}^{-2}$, $\hat{p}^{-D+2}$ and $\hat{p}_k\hat{p}_l\hat{p}^{\,-2}$
from Eqs.\ (\ref{eq:A7}), (\ref{eq:A5}) and (\ref{eq:A8}) respectively, we find that
\bea
\langle x|\Gamma_k|y\rangle=i\eta\frac{(D-2)}{(2\pi)^D}
\int d^Dz~\frac{\rlap/x-\rlap/z}{|x-z|^{D}}\,\gamma_l\,
            \frac{\rlap/z}{|z|^{D}}\,
            \frac{\partial{^y_k}\partial{^y_l}}{(\partial^2)^y}\,
            \frac{1}{|y-z|^{2}}\,.                                   \label{qed2}
\eea

Our aim is to simplify Eq.\ (\ref{qed1})  using the basic identity given in Eq.\ (\ref{key1}).
The problem in doing this is that it would lead us to (as the calculation given later
shows) applying Eq.\ (\ref{key1}) on $\hat{p}^{\,-2}\hat{q}^{\,-D}\hat{p}^{\,-D+2}$.
This (naively) results in $\hat{q}^{\,-D+2}\hat{p}^{\,-D}\hat{q}^{\,-2}$. But using Eqs.\
(\ref{eq:A5}) and (\ref{eq:A6}) for 
$\langle x|\hat{p}^{\,-D}|y\rangle$
is not possible  because $\Gamma(D/2-\alpha)$ blows up for $\alpha=D/2$. 

{\it This problem can be solved by the following
regularization of the scale dimensions}:
\bea
d_\psi=\frac{D-1-\epsilon}{2}\,,~d_A=1+\epsilon\,.
\eea
This corresponds to the propagators 
$\rlap/x/x^{D-\epsilon}\sim \rlap/p/p^{2+\epsilon}$ and
$1/x^{2+2\epsilon}\sim 1/p^{D-2-2\epsilon}$ for the fermion and the photon
respectively. The regularization of the two scale dimensions go together, since the interaction
$\bar\psi\gamma_i\psi A_i$ must continue to have the dimension $D$.
Our regularization is similar to that given in Eq.\ (2.30) of Ref.\ \cite{fradkin2},
except that we have changed the sign in front of $\epsilon$ in both $d_\psi$ and $d_A$.
This has been done to ensure that we are led to 
$\langle x| \hat{p}^{\,-D+\epsilon}|y\rangle$ in the course of our calculation
(see below), which is convergent (whereas $\langle x| \hat{p}^{\,-D-\epsilon}|y\rangle$
would diverge in the infrared).

We therefore have to simplify the regularized form of Eq.\ (\ref{qed1}):
\bea
\Gamma_k=\eta\gamma_i\gamma_l\gamma_j\hat{p}_i\hat{p}^{\,-2-\epsilon}
          \hat{q}_j\hat{q}^{\,-D+\epsilon}
          \hat{p}^{\,-D+2+2\epsilon}\hat{p}_l\hat{p}_k\hat{p}^{\,-2}\,. 
\eea
Using Eq.\ (\ref{eq:A8}) for the ``matrix element" of $\hat{p}_k\hat{p}^{\,-2}$,
we can write down 
\bea
\langle x|\Gamma_k|y\rangle=-i\eta\frac{\partial{^y_k}}{(\partial^2)^y}\langle x|\Gamma'|y\rangle\,,
                                                  \label{qed3}\\
\Gamma'=\gamma_i\gamma_l\gamma_j\hat{p}_i\hat{p}^{\,-2-\epsilon}
          \hat{q}_j\hat{q}^{\,-D+\epsilon}
          \hat{p}^{\,-D+2+2\epsilon}\hat{p}_l\,.
\eea
It is easier to put $\Gamma'$ in 
$``\hat{q}\hat{p}\hat{q}"$ form than $\Gamma_k$. First use Eq.\ (\ref{eq:A2})
to obtain
\bea
\Gamma'=\gamma_i\gamma_l\gamma_j\hat{p}_i\hat{p}^{\,-2-\epsilon}\hat{q}^{\,-D+\epsilon}
         (\hat{p}^{\,-D+2+2\epsilon}\hat{q}_j-i(D-2-2\epsilon)\hat{p}^{\,-D+2\epsilon}\hat{p}_j)
         \hat{p}_l\,.
\eea
The identity of Eq.\ (\ref{key1}) can now be used in both the terms, giving
\bea
\Gamma'=\gamma_i\gamma_l\gamma_j\hat{p}_i\hat{q}^{\,-D+2+2\epsilon}\hat{p}^{\,-D+\epsilon}
\hat{q}^{\,-2-\epsilon}\hat{q}_j\hat{p}_l
-i(D-2-2\epsilon)\gamma_i\hat{p}_i\hat{q}^{\,-D+2+2\epsilon}\hat{p}^{\,-D+\epsilon}
\hat{q}^{\,-2-\epsilon}\,.
\eea
We now bring $\hat{p}_l$ next to $\hat{p}^{\,-D+\epsilon}$ in the first term 
by moving it through $\hat{q}_j$ and then $\hat{q}^{\,-2-\epsilon}$ by using
Eqs.\ (\ref{eq:A1}) and (\ref{eq:A3}) respectively. This leads to
\bea
\Gamma'=\gamma_i\gamma_l\gamma_j\hat{p}_i\hat{q}^{\,-D+2+2\epsilon}\hat{p}^{\,-D+\epsilon}
          \hat{p}_l\hat{q}^{\,-2-\epsilon}\hat{q}_j
         +i\epsilon \gamma_i\hat{p}_i\hat{q}^{\,-D+2+2\epsilon}\hat{p}^{\,-D+\epsilon}
            \hat{q}^{\,-2-\epsilon}\,.
\eea
Finally $\hat{p}_i$ is brought next to $\hat{p}^{\,-D+\epsilon}$ in both the terms
to arrive at the $``\hat{q}\hat{p}\hat{q}"$ form:
\bea
\Gamma'&=&\gamma_i \hat{q}^{\,-D+2+2\epsilon}\hat{p}^{\,-D+2+\epsilon}
           \hat{q}^{\,-2-\epsilon}\hat{q}_i
       +i(D-2-2\epsilon)\gamma_i\gamma_l\gamma_j\hat{q}^{\,-D+2\epsilon}\hat{q}_i
        \hat{p}^{\,-D+\epsilon}\hat{p}_l\hat{q}^{\,-2-\epsilon}\hat{q}_j    \nonumber\\
       &&+i\epsilon \gamma_i\hat{q}^{\,-D+2+2\epsilon}\hat{p}^{\,-D+\epsilon}
         \hat{p}_i\hat{q}^{\,-2-\epsilon}
       -\epsilon(D-2-2\epsilon)\gamma_i\hat{q}^{\,-D+2\epsilon}\hat{q}_i\hat{p}^{\,-D+\epsilon}
        \hat{q}^{\,-2-\epsilon}\,.                                       \label{four}
\eea
The evaluation of $\langle x|\Gamma'|y\rangle$ can now be completed by using Eqs.\ 
(\ref{eq:A4})-(\ref{eq:A7}). It is found that in the resulting terms, the diverging
$\Gamma(\epsilon/2)$ always comes multiplied by $\epsilon$. Since
$\epsilon\Gamma(\epsilon/2)=2\Gamma(1+\epsilon/2)$, taking $\epsilon\rightarrow 0$
gives finite results for all the four terms of Eq.\ (\ref{four}) 
(with the third term giving zero). We then obtain
\bea
\langle x|\Gamma'|y\rangle&=&\frac{1}{\pi^{D/2}2^{D-2}\Gamma(D/2-1)}\,
              \frac{x^2\rlap/y-\rlap/x(\rlap/x-\rlap/y)\rlap/y-\rlap/x |x-y|^2}
                   {x^D |x-y|^2 |y|^2}\\
             &=&\frac{1}{\pi^{D/2}2^{D-2}\Gamma(D/2-1)}\,\frac{\rlap/x}{x^D}
                \Bigg(\frac{1}{|x-y|^2}-\frac{1}{|y|^2}\Bigg)\,.         \label{qedx}
\eea
Since $(\partial^2)^y \ln(|x-y|/|y|)=(D-2)(1/|x-y|^2-1/|y|^2)$,
Eqs.\ (\ref{qed3}) and  (\ref{qedx}) lead to
\bea
\langle x| \Gamma_k|y\rangle=-i\eta\frac{1}{(4\pi)^{D/2}\Gamma(D/2)}\,\frac{\rlap/x}{x^D}
                 \partial{^y_k}\ln\frac{|x-y|^2}{|y|^2}\,.                  \label{qedy}
\eea
The right hand sides of Eqs.\ (\ref{qed2}) and (\ref{qedy}) are now to be equated.
In terms of the variables $x_1$, $x_2$, $x_3$ and $x_4$ defined 
below Eq.\ (\ref{key2}), the resulting relation reads
\bea
\int d^Dx_4~ \frac{\rlap/x_{14}}{x_{14}^D}\,\gamma_l\,
              \frac{\rlap/x_{42}}{x_{24}^D}\,
              \frac{\partial{^{x_3}_k}\partial{^{x_3}_l}}{(\partial^2)^{x_3}}\,
               \frac{1}{x_{34}^2}\,
         &=&\frac{\pi^{D/2}}{(D-2)\Gamma(D/2)}\,\frac{\rlap/x_{12}}{x_{12}^D}\,
            \partial{^{x_3}_k}\ln\frac{x_{23}^2}{x_{13}^2}        \label{qedz1}     \\
         &=&\frac{2\pi^{D/2}}{(D-2)\Gamma(D/2)}\,\frac{\rlap/x_{12}}{x_{12}^D}\,
           \Bigg(\frac{(x_{13})_k}{x_{13}^2}-\frac{(x_{23})_k}{x_{23}^2}\Bigg)\,.
                                                            \label{qedz2}
\eea
Eqs.\ (\ref{qedz1}) and (\ref{qedz2}) agree with the longitudinal structure function given 
from general considerations of conformal invariance in 
Refs.\ \cite{palchik} and \cite{comm} respectively
(the fermion scale dimension being $d_\psi=(D-1)/2$ in our case).
From Eq.\ (\ref{qedz2}), we note the value of the coefficient for the physically interesting
cases: $\pi^2$ for $D=4$ (massless QED$_4$)
and $4\pi$ for $D=3$ (massless QED$_3$ in the infrared).
\section{Conclusion} 
\label{concl}
In this work, we evaluated conformal invariant integrals
involving spin one-half and spin-one particles in the context of two
$D$-dimensional field theories with tree-level scale invariance:
the massless Yukawa theory and massless QED, both with dimensionless
coupling constants. 
The three-point function of the Yukawa theory
and the longitudinal part of the three-point function of QED
were explicitly evaluated to the lowest order, and the results were
expressed in conformal invariant forms. We made use of the operator algebraic method of
calculating  massless Feynman integrals. For the QED calculation,
regularization of the scale dimensions of the particles was used. 
While the present work focused on the longitudinal part only, our plan is to
evaluate the entire
QED three-point function to the lowest order.
The result can then be used in higher order studies of massless QED$_3$ in the infrared,
and also for implementing the bootstrap program in that theory.
More generally, the techniques developed in the present work
should be useful for calculations in other 
massless field theories and $D$-dimensional CFT's.
\section*{Acknowledgements}
I am indebted to H. S. Sharatchandra for introducing me to the
topics of this paper. I also
thank S. D. Joglekar for discussions.
\appendix
\section*{Appendix}
In this Appendix, we list and develop some important formulas of the operator approach
to the evaluation of  massless Feynman integrals.
We use $i,j,k,\cdots$ for spacetime indices, and $\alpha,\beta,\cdots$ for exponents of
$\hat{q}^{\,2}$ and $\hat{p}^{\,2}$. Thus,
$\hat{q}^{\,2\alpha}=(\sum_i \hat{q}_i \hat{q}_i)^\alpha$ (and likewise
$\hat{p}^{\,2\alpha}$), the parameter $\alpha$ being in general a complex 
number \cite{isaev1}.
The fundamental commutation relation
\bea
[\hat{q}_i, \hat{p}_j]=i\delta_{ij}                              \label{eq:A1}
\eea
leads to the following two useful relations:
\bea
[\hat{q}_i,\hat{p}^{\,2\alpha}]=i2\alpha\hat{p}^{\,2\alpha-2}\hat{p}_i\,, \label{eq:A2}\\
\,[\hat{p}_i,\hat{q}^{\,2\alpha}]=-i2\alpha\hat{q}^{\,2\alpha-2}\hat{q}_i\,. \label{eq:A3}
\eea
(A check on Eqs.\ (\ref{eq:A2}) and (\ref{eq:A3}) is that they immediately give us
Eqs.\ (13) and (14) of Ref.\ \cite{isaev1}.)

We use the normalization of position and momentum eigenstates followed in Ref.\
\cite{isaev1}. This results in the following two 
``matrix elements" \cite{isaev1}
\bea
\langle x|\hat{q}^{\,2\alpha}|y\rangle=|x|^{2\alpha}\delta^{(D)}(x-y)\,,   \label{eq:A4}\\
\langle x|\hat{p}^{\,-2\alpha}|y\rangle=a(\alpha)\frac{1}{|x-y|^{D-2\alpha}}\,,  \label{eq:A5}
\eea
where
\bea
a(\alpha)=\frac{\Gamma(D/2-\alpha)}{\pi^{D/2}2^{2\alpha}\Gamma(\alpha)}\,. \label{eq:A6}
\eea
In Eq.\ (\ref{eq:A5}), $D/2-\alpha\neq 0, -1, -2, \cdots$. Now,
$\langle x|\hat{p}_i\hat{p}^{\,-2\alpha}|y\rangle=-i\partial{^x_i}\langle x|\hat{p}^{\,-2\alpha}|y\rangle$
(this being obtained by inserting the
completeness relation in momentum space on the left-hand side).
Eq.\ (\ref{eq:A5}) then gives us
\bea
\langle x|\hat{p}_i\hat{p}^{\,-2\alpha}|y\rangle=i(D-2\alpha)
         a(\alpha)\frac{(x-y)_i}{|x-y|^{D-2\alpha+2}}\,.        \label{eq:A7}
\eea
Another useful ``matrix element" which can be similarly obtained is
$\langle x|\hat{p}_i|y\rangle=i\partial{^y_i}\delta^{(D)}(x-y)$.
This relation can be generalized to
\bea
\langle x|f(\hat{p}_i)|y\rangle=f(i\partial{^y_i})\delta^{(D)}(x-y)           \label{eq:A8}
\eea
where $f$ denotes an arbitrary function. As a check, it may be noted that consistency
of Eq.\ (\ref{eq:A8}) with Eqs.\ (\ref{eq:A5}) and (\ref{eq:A6})
lead to the expression for the Green function for the operator
$((-\partial^2)^y)^\alpha$.

\end{document}